\documentclass[conference]{IEEEtran}
\IEEEoverridecommandlockouts
\usepackage{cite}
\usepackage{amsmath,amssymb,amsfonts}
\usepackage{algorithmic}
\usepackage{graphicx}
\usepackage{textcomp}
\usepackage{xcolor}
\usepackage{hyperref}
\def\BibTeX{{\rm B\kern-.05em{\sc i\kern-.025em b}\kern-.08em
    T\kern-.1667em\lower.7ex\hbox{E}\kern-.125emX}}
\begin{document}

\title{RATE: An LLM-Powered Retrieval Augmented Generation Technology-Extraction Pipeline \\

}

\author{\IEEEauthorblockN{1\textsuperscript{st} Karan Mirhosseini\textsuperscript{ *}\thanks{* Corresponding author}}
\IEEEauthorblockA{\textit{\small Dept. of Management, Science and Technology} \\
\textit{Amirkabir University of Technology}\\
Tehran, Iran \\
karan@aut.ac.ir}
\and
\IEEEauthorblockN{2\textsuperscript{nd} Arya Aftab}
\IEEEauthorblockA{\textit{\small Dept. of Electrical Engineering} \\
\textit{Sharif University of Technology}\\
Tehran, Iran \\
arya.aftab@alum.sharif.edu}
\and
\IEEEauthorblockN{3\textsuperscript{rd} Alireza Sheikh}
\IEEEauthorblockA{\textit{\small Dept. of Management, Science and Technology} \\
\textit{Amirkabir University of Technology}\\
Tehran, Iran \\
a.sheikh@aut.ac.ir}
}

\maketitle

\begin{abstract}

In an era of radical technology transformations, technology maps play a crucial role in enhancing decision making. These maps heavily rely on automated methods of technology extraction. This paper introduces Retrieval Augmented Technology Extraction (RATE), a Large Language Model (LLM) based pipeline for automated technology extraction from scientific literature. RATE combines Retrieval Augmented Generation (RAG) with multi-definition LLM-based validation. This hybrid method results in high recall in candidate generation alongside with high precision in candidate filtering. While the pipeline is designed to be general and widely applicable, we demonstrate its use on 678 research articles focused on Brain-Computer Interfaces (BCIs) and Extended Reality (XR) as a case study. Consequently, The validated technology terms by RATE were mapped into a co-occurrence network, revealing thematic clusters and structural features of the research landscape. For the purpose of evaluation, a gold standard dataset of  technologies in 70 selected random articles had been curated by the experts. In addition, a technology extraction model based on Bidirectional Encoder Representations of Transformers (BERT) was used as a comparative method. RATE achieved F1-score of 91.27\%, Significantly outperforming BERT with F1-score of 53.73\%.  Our findings highlight the promise of definition-driven LLM methods for technology extraction and mapping. They also offer new insights into emerging trends within the BCI-XR field. The source code is available {\color{purple}\url{https://github.com/AryaAftab/RATE}}
\end{abstract}


\section{Introduction and Literature Review}
The early 21st century has been an era of acceleration in the development of transformative digital technologies. Powered by advancements in computing power, high-resolution displays, and artificial intelligence (AI), new ways of experiencing reality have emerged, most notably Augmented Reality (AR) and Virtual Reality (VR) \cite{rauschnabel2022xr}. AR enhances users' perception of their real world environment by overlaying digital data onto their view \cite{dargan2023augmented}, while VR immerses users in a completely artificial, computer generated reality \cite{rauschnabel2022xr}, the umbrella term that brings these two immersive technologies together is called Extended Reality (XR) \cite{kohli2022review}.

These immersive tools have found potent applications across diverse fields. In medicine, they help with diagnostics, surgical procedures, and rehabilitation programs \cite{yeung2021virtual}. Educational settings benefit from these technologies for training, interactive teaching, and observational learning \cite{motejlek2021taxonomy}. Furthermore, in industrial contexts, AR and VR serve as critical tools for maintenance, assembly guidance, and improving human-robot collaboration \cite{reljic2021augmented}. The efficacy of these applications often stems from the realistic and engaging experiences these technologies provide. Such experiences can be significantly broadened and enriched by combining XR with Brain-Computer Interface (BCI) \cite{putze2020brain}. A BCI functions as a communication system that allows direct control of computers or external devices through neural activity alone \cite{nicolas2012brain}. In AR, BCIs offer a powerful method for real-time engagement with AR headsets, leading to applications such as hands-free manipulation of information by medical professionals, direct neural command of robotic devices, and help with industrial or smart environment controls. Furthermore, within VR systems, this enhanced BCI connectivity facilitates the creation of deeply immersive artificial environments. Such capabilities are fundamental to a variety of VR applications, including thought-based navigation through virtual spaces, entertainment experiences dynamically tailored to a user's cognitive or emotional state, and comprehensive rehabilitation programs where immersive scenarios help patients engage in therapeutic exercises \cite{kohli2022review}.

Despite significant potentials and diverse applications, the integration of BCI-XR systems still faces several challenges. One of the most important constraints is at the hardware and technology level, as XR and BCI devices worn simultaneously can introduce mutual interference, leading to malfunction \cite{si2017brain}. This integration also introduces serious ethical challenges, notably regarding user privacy, data security, and adverse cognitive consequences \cite{piszcz2024impact}. These technical and ethical complexities, along with the rapidly evolving BCI-XR landscape, require a comprehensive approach that systematically maps and plans policies to improve their societal impacts \cite{sun2016mapping}. One of the main steps of decision making regarding such technologies is "\textit{Technology Identification}", defined as the process of developing a list of technologies which are or may be present in the field, firm, or product. Many methods for the latter have been developed. However, when the scope of the identification is the whole field and not just a product or a firm, "\textit{Technology Mapping}" has proved to be the best method \cite{arasti2010use}.

Technology maps are visual representations of the core and peripheral technologies and their dynamics within a field. A well-established technique for constructing such maps and defining technologies and their relationships involves analyzing the co-word occurrence of technology terms within the abstracts, keywords, and titles of scientific articles of the field \cite{castells2000technology}. However, the successful creation of these informative co-occurrence networks fundamentally depends on the accurate and consistent extraction of specific technology terms from vast amounts of contextual data. Due to the significant growth in scientific literature, manually reviewing these documents to identify specific technology terms is no longer feasible. Therefore, an automated approach is required \cite{firoozeh2020keyword}. Key methods to automating this task include: Rule-based linguistic methods which benefit from deep linguistic knowledge, yielding potentially high accuracy. However, they are computationally demanding and require significant domain expertise. Statistical techniques, on the other hand, utilize features from large corpora, offering language independence and achieving good results with sufficient data. Nonetheless, sometimes with less precision than linguistic methods. Finally, machine learning strategies, often supervised, learn from tagged training documents to identify relevant terms in new texts \cite{siddiqi2015keyword}. In one of the most recent approaches Puccetti et al. suggested a BERT-based method in combination with a rule-based method. Combining the machine-learning method with rule-based models they achieved 40 percent precision for the task of technology entity recognition. The main issue addressed by the researchers causing the low precision was lack of annotated datasets for technologies within different fields \cite{puccetti2023technology}.

The critical need for task-specific training via annotated data can be addressed by the use of Large Language Models (LLMs). While Pre-trained Language Models (PLMs) like BERT introduced task-agnostic pre-training on text with subsequent fine-tuning, LLMs represent a significant advancement. These exceptionally scaled, transformer-based networks, trained on massive corpora, possess superior language understanding and crucial emergent abilities not found in smaller PLMs. Such capabilities, including in-context learning from prompts, instruction following, and multi-step reasoning, can reduce the critical need for extensive task-specific training on annotated data for many applications \cite{minaee2025largelanguagemodelssurvey,zhao2023survey}. Thanks to recent advancements, LLMs can tackle novel tasks without specific training (zero-shot), leveraging their pre-trained general knowledge via in-context learning provided in prompts \cite{gadetsky2025large}. Nonetheless, these models encounter critical issues such as hallucination and outdated data especially in knowledge intensive tasks. Addressing these issues, Retrieval Augmented Generation (RAG) had been introduced. RAG systems retrieve relevant documents to the task from an external database based on similarity calculations, a process which enhances the performance of LLMs for special tasks significantly \cite{gao2023retrieval}.

Challenges with precision in prior automated technology recognition methods, due to the lack of annotated datasets, highlight the need for a more reliable and generalizable approach. Although modern LLMs and RAG systems precisely execute tasks, they do not inherently guarantee that extracted terms genuinely represent "\textit{Technology}" without validation against scholarly criteria. The crucial missing element is a targeted methodological framework for this rigorous definitional assessment. Therefore, a significant methodological gap exists for a systematic approach.

Addressing this identified methodological gap, this paper introduces Retrieval Augmented Generation Technology Extraction (RATE), an automated pipeline designed for general-scale technology extraction from textual data. RATE employs a distinctive multi-stage process: it first utilizes an LLM powered by RAG for comprehensive candidate technology identification from scientific texts (abstracts, titles, and keywords). These candidates then undergo a novel validation phase where a second LLM critically evaluates each term against four distinct, literature-derived definitions of technology, considering the full textual context from which the candidate originated. This study applies the RATE pipeline to a corpus of scientific literature focused on the intersection of AR, VR, and BCI. Our primary aims are to demonstrate the feasibility and effectiveness of this definition-driven, LLM-powered methodology in creating a granular and rigorously validated inventory of technologies, and subsequently, to provide initial insights into the technological landscape of the BCI-XR domain through co-occurrence analysis.

\section{Methodology}

\subsection{Introduction to Methodology}

\begin{figure*}[t]
\vspace{-1em}
			\centering
			\includegraphics[width=175mm,scale=1.0]{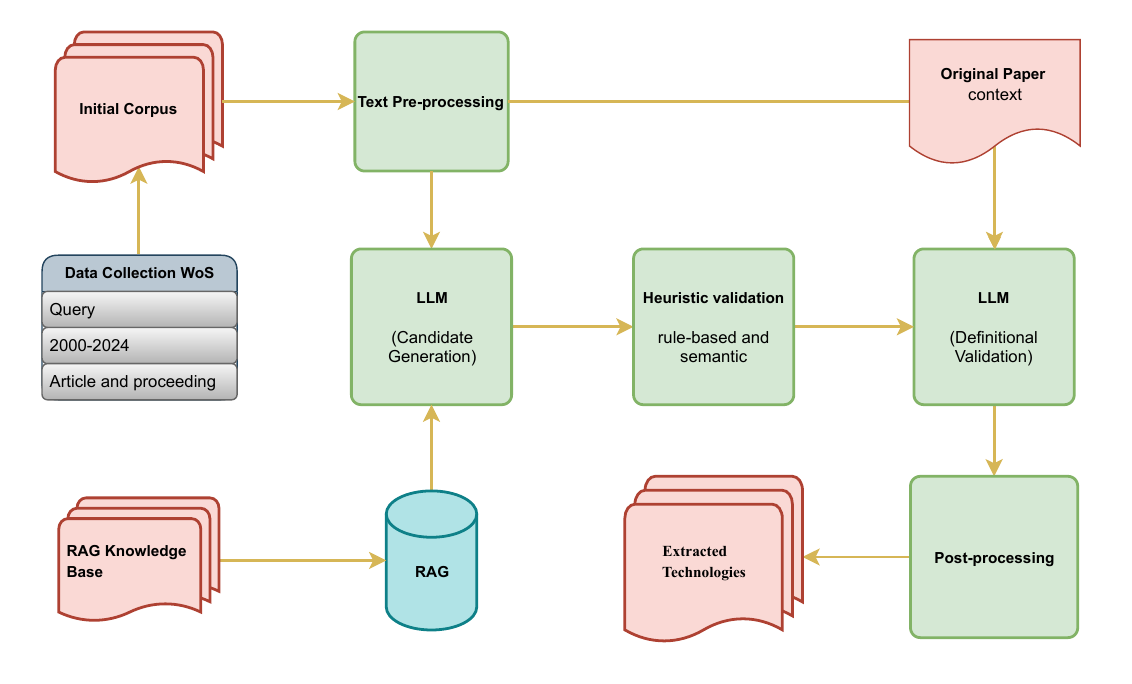}
				\caption{\small An overview of technology extraction and validation pipeline}
				\label{fig:scheme}
			\end{figure*}

This study aims to develop and evaluate RATE, an automated pipeline for extracting and validating technology terms from scientific literature, specifically focusing on publications at the intersection of AR, VR, and BCI. The proposed multi-stage methodology integrates LLMs with RAG for initial candidate identification, followed by a novel LLM-based validation process employing multiple scholarly definitions of technology. This approach was chosen for its potential to achieve high recall in candidate generation, enhance contextual understanding through RAG, and ensure high precision and conceptual rigor in the final validated list through its unique definitional validation step, addressing limitations of prior automated methods (Fig~\ref{fig:scheme}).

\subsection{Data Collection and Preprocessing}

The primary corpus for this study consists of data gathered from the Web of Science (WoS) Core Collection. The search query was constructed in two main parts: the first part focused on AR and VR terms, and the second focused on BCI terms. These two parts were then combined using the 'AND' operator of WoS to retrieve publications at the intersection of these fields.

The search string for AR and VR, based on the work of Yeung et al. \cite{yeung2021virtual}, was: TS= ("\textit{virtual realit}*" OR "\textit{augmented realit}*" OR "\textit{mixed realit}*" OR "\textit{computer-mediated realit}*")

The search string for BCI, based on the work of Yin et al. \cite{yin2023bibliometric}, was: TS= ("\textit{brain-computer interface}*" OR "\textit{brain-machine interface}*" OR "\textit{brain machine interface}*" OR "\textit{brain computer interface}*" OR "\textit{direct neural interface}*")

Therefore, the final combined search string used in WoS was: TS= (("\textit{virtual realit}*" OR "\textit{augmented realit}*" OR "\textit{mixed realit}*" OR "\textit{computer-mediated realit}*") AND ("\textit{brain-computer interface}*" OR "\textit{brain-machine interface}*" OR "\textit{brain machine interface}*" OR "\textit{brain computer interface}*" OR "\textit{direct neural interface}*"))

To refine the search results, the publication years were restricted to 2000–2024, and document types were limited to  "\textit{proceeding}" and "\textit{article}". Furthermore, the language was set to English only. This search strategy yielded 678 publications as of the 23rd of May 2025. For each of these publications bibliographic data including Abstract, Title, publication year, and author keywords were extracted. The extracted Title, Abstract, and Author Keywords for each publication were then concatenated to form a single comprehensive text document representing that publication. This combined text served as the primary input for the subsequent technology extraction process. In the final pre-processing step a dedicated function was used to standardize the format of the text document by removing excessive whitespaces.

\subsection{RATE Pipeline}

\subsubsection{\textbf{\small RAG Setup}}

The initial stage of the RATE pipeline establishes a RAG system designed to provide the primary technology extraction LLM with relevant contextual documents. resulting in domain understanding and term disambiguation for the LLM. RAG's knowledge base is  constructed by consolidating diverse public technology lists, including Wikipedia’s list of emerging technologies (drawing on insights from Puccetti et al. \cite{puccetti2023technology}), the Chinese catalogue of prohibited and restricted technologies (based on CSET \cite{CSET2023}), the International Energy Agency’s (IEA) list of clean energy technologies (based on IEA’s ETP clean energy technology \cite{IEA2025}) and O*NET’s list of technology skills and tools (based on National Center for O*NET Development \cite{ONET2024}). Entries from these sources were structured into documents with standardized fields  of \textit{name}, \textit{type}, \textit{domain}, and \textit{description}. These documents were then segmented into manageable chunks using \textit{RecursiveCharacterTextSplitter} of \textit{LangChain} library \cite{Chase2024} and subsequently vectorized using the \textit{OllamaEmbeddings} model \textit{mxbai-embed-large} \cite{Ollama2025,emb2024mxbai}. The resulting standardized numerical embeddings, along with their corresponding text chunks and metadata, are then stored locally. From this store, a retriever is configured to fetch an initial set of relevant documents based on distance (the top 20 most similar) for any given input (\textit{abstract}, \textit{keyword}, and \textit{title}). Finally, this retrieved set undergoes a custom, two-pass diversity filtering process. This process first prioritizes documents exhibiting unique metadata combinations (specifically \textit{type} and \textit{domain} fields) and subsequently fills the selection with additional documents from the retrieved set if needed, aiming for a target of up to seven diverse documents. This method ensures a varied yet relevant contextual input for the technology extraction LLM.

\subsubsection{\textbf{\small LLM-based Candidate Technology Extraction}} \label{subsubsec:Candidate_Extraction}

For the initial candidate technology extraction, the RATE pipeline utilizes the \textit{DeepSeek-V3} model, accessed via its API endpoint \cite{liu2024deepseek}. Key generation parameters were set to ensure deterministic and comprehensive output, including a \textit{temperature} of 0.0 and \textit{max\_tokens} set to 4096. This model is then prompted using a precisely engineered strategy for technology extraction with the objective of achieving maximum recall. It cautiously analyzes the primary input text to extract every single term or phrase that even remotely suggests a technology. Within this process, the LLM assigns each candidate a confidence score and includes the term if this score meets or exceeds a threshold of 0.7. While diverse contextual documents are provided via the RAG system, they serve strictly for background understanding and disambiguation. As a result, all extracted technologies originate from the primary input text. This process aims to ensure that a comprehensive initial set of potential technologies are captured for subsequent refinement.

\subsubsection{\textbf{\small Intermediate Heuristic Validation}}

Following the initial candidate generation by the LLM, the RATE pipeline incorporates an intermediate heuristic validation stage. first, a candidate is validated if its full phrase (case-insensitive), its base term (if the candidate contains a parenthesized acronym), or its potential acronym (if derivable from the candidate) is directly found within the input text. Second, for multi-word candidates, a partial compound match is accepted if a significant proportion ($\geq 0.75$) of its constituent meaningful words appear in the source text. Furthermore, candidates are provisionally retained if they were assigned a very high confidence score ($\geq 0.95$) by the initial LLM. As a final validation measure for candidates with at least moderate initial confidence ($\geq 0.75$) that did not meet the preceding criteria, their semantic similarity to the entire input text is calculated using a \textit{spaCy} model. if this similarity exceeds a predefined threshold ($\geq 0.70$), the term is included. All these steps were designed to catch the probable hallucinations of the LLM from the first step. Proceeding from this stage, a refined, yet still broad, list of candidate phrases has been created. Yet their status as ‘technology’ needs to be validated. Therefore, a novel LLM-based validation method was subsequently employed for this critical technology identification task.

\subsubsection{\textbf{\small LLM-based Definitional Validation}}

The fourth stage of the RATE pipeline subjects the output list of the previous step to a definitional assessment using a second LLM instance. \textit{DeepSeek-V3} with the exact same settings as the previous step was incorporated in this step. For each candidate phrase, LLM was provided with both the phrase itself and the complete original input text to ensure a comprehensive contextual evaluation. Central to this validation is a detailed system prompt that instructs LLM to act as an expert technology analyst. Its core task is to precisely evaluate whether the candidate, viewed within its specific textual context, qualifies as a technology or not. To enhance the understanding of the LLM from technology, four distinct, pre-defined scholarly definitions of technology based on the works of Puccetti et al. that were explicitly provided in its instructions \cite{puccetti2023technology}. The prompt further guides the LLM to focus its interpretation of these definitions on tangible, applied technical implementations, tools, systems, or methods, using specific 'YES if' / 'NO if' decision criteria. For every candidate, the LLM is required to return its assessment in a structured \textit{YAML} format, which includes a Boolean field, indicating its textual reasoning for its decision, an integer confidence from 1 to 10 (based on detailed scoring rubrics provided to the LLM), and the technology name. A candidate technology phrase is ultimately confirmed and retained from this stage if the LLM returns Boolean field: true and assigns a confidence score greater than six. This step enhances the precision of the model significantly while maintaining the recall. While the results of this step are close to the final list of technologies within the specific paper, due to the nature of LLMs output some redundancies and different letter case variances for phrases were observed. This issue was addressed in the next step.

\subsubsection{\textbf{\small Post-Processing}}

Following the LLM-based definitional validation, the confirmed technology terms undergo a final series of post-processing steps. First, all validated terms are converted to lowercase, and exact duplicates are removed to ensure a unique set. This list is then checked against a user-defined, domain-specific list of prohibited or overly generic terms, allowing for user-defined exclusion of technologies. Subsequently, an acronym-priority deduplication was performed to manage instances where both an acronym and its full expanded form exists, retaining the more concise version if they refer to the same base concept. The final remaining technology terms are then sorted alphabetically, yielding the refined list of technologies extracted from each paper.

\subsection{Output Evaluation}

To evaluate the performance of the RATE pipeline, a ground truth dataset was established. First, 70 scientific papers were randomly selected from the 678 collected documents. Then, three domain experts independently reviewed the title, abstract, and author keywords for each of these papers to manually compile a gold-standard list of technologies. In the next step, same technologies extracted by all experts were added to the final list and the remaining technologies were debated whether to be omitted or to be added to the final list. The result was a final gold standard list approved by all experts. Following this, the list of technologies extracted by the RATE pipeline for these same 70 papers was presented to each expert. They were then tasked with identifying True Positives (TP), False Positives (FP), and False Negatives (FN) by comparing RATE's output against the final curated gold-standard list. Based on experts' assessment, Precision and Recall were calculated. In the next step the comparison of the proposed model with another state-of-the-art model was required. Therefore, a supervised machine learning model has been trained for the task of the technology extraction from text so its results would be compared with RATE’s output and the gold standard list of technologies.

\subsection{BERT baseline for technology extraction}

Bidirectional Encoder Representations from Transformers (BERT) is a pretrained transformer with ability to achieve exceptional results in many tasks including sequence classification \cite{devlin2019bert, liu2021ner}. Using this model, we aimed to imply a supervised machine learning model for the technology extraction task which then enabled us to quantify the precision and F1-score of our model which did not benefit from a training phase.

The BERT model was fine-tuned on a custom NER dataset comprising approximately 2200 rows of tokenized scientific sentences from domains like neuroscience, robotics, and VR/AR, sourced from open-source publications. BIO tags were used in the task of curating this dataset, \textit{B-TECH} indicating the beginning of a technology, \textit{I-TECH} for inside of a technology entity, and \textit{O} for a non-technology token. The dataset were split into training (80\%), validation (10\%), and test (10\%) sets.

"\textit{bert-base-cased}" variant from \textit{Hugging Face Transformers} library were used as the base model, initialized with 3 output labels corresponding to the BIO tags \cite{wolf2020huggingfacestransformersstateoftheartnatural}. The dataset was preprocessed into a custom NERdataset class, handling tokenization, padding/truncation to a maximum length of 128 tokens, and alignment of labels with subword tokens. Training was conducted using the Trainer API with the following hyperparameters: 20 epochs, batch size of 16, learning rate of 2e-5, AdamW optimizer with weight decay of 0.01, and early stopping based on F1-score on the validation set. Evaluation metrics included token-level accuracy and entity-level F1-score, accounting for partial matches in multi-word entities. This BERT baseline was trained on a GPU-enabled environment, achieving convergence after 15 epochs. Post-training, the model was used to predict technology entities on unseen test texts by aggregating \textit{B-TECH} and \textit{I-TECH} tokens into phrases, filtering out single-token noise through post-processing.

\subsection{Technology Mapping and Co-occurrence Analysis}

Following the extraction and validation of technology terms by the RATE pipeline for each of the 678 scientific papers, a co-occurrence analysis was conducted to map and visualize the relational structure of these technologies within the BCI-XR research domain. This analysis aimed to identify prominent technologies, understand their interconnections, and reveal underlying thematic groupings. Two technologies were considered to co-occur if they both appeared in the final validated list of extracted technologies for the same scientific paper. This co-occurrence data was used to construct a network where each unique validated technology represents a node, and an edge between two nodes signifies that the technologies co-occurred. The weight of each edge was determined by the frequency of co-occurrence across the entire corpus of 678 papers. The co-occurrence network was constructed using custom Python scripts leveraging the \textit{Pandas} library \cite{reback2020pandas} for reading the processed Excel data and the \textit{NetworkX} library for graph creation and manipulation \cite{hagberg2008exploring}. Nodes with low edge wight were removed ($<10$) for better readability, then the average weighted degree, indicating the average sum of co-occurrence frequencies for each technology node, was calculated. The network diameter, defined as the longest shortest path between any two nodes, was determined to assess the overall compactness of the technology landscape. Furthermore, to evaluate the presence and strength of community structure within the network, modularity was calculated using the Louvain method \cite{blondel2008fast} and Girvan-Newman method \cite{newman2004finding} as implemented in \textit{Gephi} \cite{bastian2009gephi}. Finally, the average clustering coefficient, which measures the local cohesiveness and the propensity for technologies to form tightly-knit groups, was computed. These metrics were selected to provide a comprehensive quantitative description of the network's topology, connectivity, and community organization. visual representation and exploration of the technology network maps, including the depiction of these communities and overall structural properties, \textit{VOSviewer} was primarily utilized, employing its specific layout and visualization algorithms \cite{van2009software}. These quantitative metrics, alongside the visual maps, were selected to provide a comprehensive description of the network's topology, connectivity, and community organization.  the specific values and their implications for the BCI-XR field are presented in the Results and Discussion sections.

\subsection{Methodological limitations}

Inherent biases in LLMs resulting from their pre-training phase are critical aspects requiring careful consideration \cite{li2025understanding}, in sensitive fields like medicine, such biases could potentially lead to adverse outcomes for patients \cite{ayoub2024inherent}. In our work these biases could lead to misidentification of technologies. Furthermore, while Web of Science is a well-regarded tool for scientific data collection, it, like any database, possesses inherent biases and limitations, and a consensus on an optimal universal source for publication data has not yet been established \cite{pranckute2021web}. A major methodological limitation encountered during our experiments was the sensitivity of the initial LLM processing stage(~\ref{subsubsec:Candidate_Extraction}) to prompt design. Even with efforts to employ deterministic settings, the outputs varied significantly depending on the specific prompts utilized. Consequently, using LLMs as a tool for entity extraction demands surgical precision in prompt engineering, which can be time-consuming and complex. Another significant challenge related to the contested nature of the term technology itself, many historical definitions are heavily debated, underscoring the difficulty of applying a single, universally accepted definition in automated tasks \cite{agar2020technology}. This lack of a universal definition posed considerable complexities specifically in curating the RAG knowledge base and configuring the second LLM's validation process. Moreover, while RATE does not incorporate domain-specific elements, its optimal performance and viability across diverse scientific disciplines would necessitate further extensive experimentation. Finally, the multi-stage approach designed to maintain high levels of both recall and precision, inherently involves significant complexity and computational expense.

\begin{table*}[t]
\caption{Network statistics overview}
\centering
\begin{tabular}{|l|c|p{9cm}|}
\hline
\textbf{Parameter} & \textbf{Value} & \textbf{Indication} \\
\hline
Number of nodes & 1260 & Total number of unique technologies identified and included in the network. \\
Number of edges & 11610 & Total number of co-occurrence relationships occurred. \\
Network diameter & 3 & The longest shortest path between any two technologies in the network. \\
Network density & 0.015 & Ratio of present edges to the maximum possible edges. \\
Clustering co-efficient & 0.912 & The degree to which technologies in the network tend to cluster together. \\
Average path length & 2.066 & The mean of shortest-path distance between all pairs of nodes. \\
Average weighted degree & 23.041 & The average sum of the weights of edges connected to a node. \\
Girvan-Newman modularity & 0.44 & A measure of the strength of division of a network into communities. \\
Louvain modularity & 0.395 & A measure of the strength of division of a network into communities. \\

\hline
\end{tabular}
\label{tab:network_info}
\end{table*}

\section{Results}

\subsection{Performance Evaluation of RATE and BERT}

To evaluate the performance of the RATE pipeline, three gold-standard labeled lists were created by three domain experts who independently reviewed 70 randomly selected papers from the primary corpus. For each of these 70 papers, the experts manually identified technologies based on the combined title, abstract, and author keywords, this identification and selection of technologies were based on the same definitions presented to the RATE. After discussing the potential terms indicating technologies, a single gold standard list based on the previous three were created. The technologies extracted by RATE and BERT for these 70 papers were then compared against the gold-standard list to identify True Positives (TP), False Positives (FP), and False Negatives (FN) of both models. Consequently, precision, recall, and F1-score were calculated for RATE and BERT.

Calculated precision, recall, and F1-score for BERT were 39.16\%, 85.50\%, and 53.73\% respectively. While the results for RATE were significantly superior with 94.26\% precision, 88.47\% recall, and 91.27\% F1-score. RATE's superior F1-score highlights its effectiveness in balancing high precision with strong recall, without requiring annotated training data. To the best of our knowledge, this F1-score represents a state-of-the-art performance for technology extraction in a general context from scientific literature, particularly in zero-shot settings.

\subsection{Descriptive Statistics of Extracted Technologies}

The RATE pipeline was applied to the entire corpus of 678 scientific papers focusing on the intersection of XR and BCI, published between 2000 and 2024. This comprehensive analysis identified and validated a total of 3,181 unique technology terms across all documents. Regarding the distribution of technologies per paper, the most frequently occurring counts of unique technologies were 8 (observed in 78 papers), 7 (in 76 papers), and 9 (in 75 papers). On average, each paper contained 8.8 unique technology terms, with a standard deviation of 3.4. The number of unique technologies identified in a single paper ranged from a minimum of 2 to a maximum of 23. Analysis of corpus frequency revealed that the most widely observed technologies were BCI at 71.64\%, VR at 50.66\%, Electroencephalography (EEG) at 43.13\%, and AR at 17.73\%. Other notable frequently occurring technologies included Motor Imagery (MI) (14.18\% corpus frequency), Steady State Visual Evoked Potential (SSVEP) (6.06\%), Machine Learning (ML) (5.32\%), Neurofeedback (5.17\%), eye tracking technologies (3.84\%), P300 (2.95\%), and Head-Mounted Displays (HMDs) (2.66\%).

\subsection{Technology mapping using co-occurrence networks}

\begin{figure}[t]
			\centering
			\includegraphics[trim={0cm -1.5cm 0cm -1.5cm}, clip, scale = 0.22]{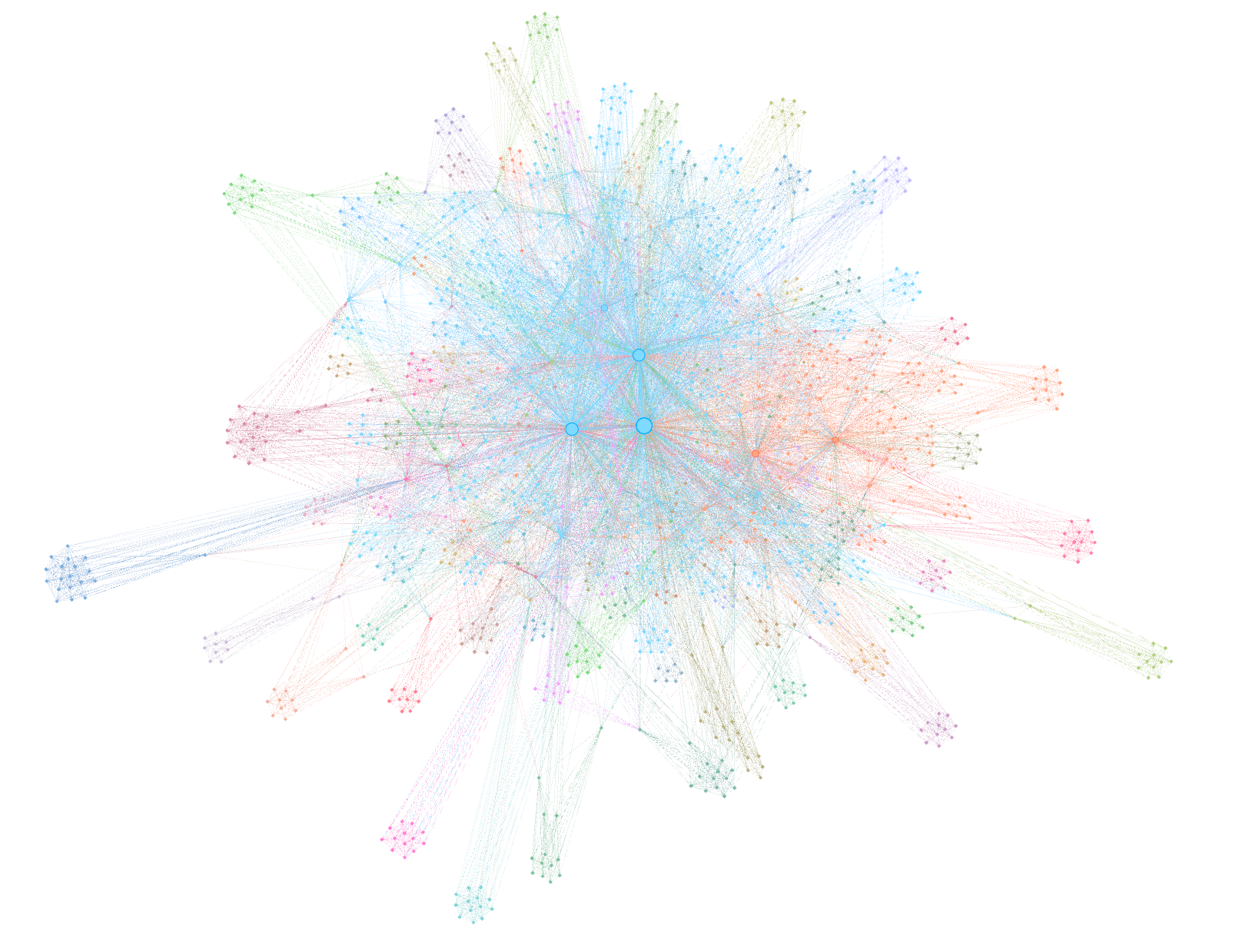}
				\caption{\small Representation of Girvan-Newman communities in the network}
				\label{fig:full_girvan}
				
			\vspace{-1mm}
			\end{figure}

The resulting technology co-occurrence network, initially consisted of 3181 nodes (representing unique technologies) and 23465 edges (representing co-occurrence of technologies). However, analyzing and visualizing such a large network typically results in a confusing representation. Therefore, to improve clarity and focus on more significant relationships, nodes with a weighted degree of less than 10 were removed. This filtering process resulted in a final primary network of 1260 nodes and 11610 edges (Table~\ref{tab:network_info}). This refined network were then created and analyzed with the help of Gephi, a prominent network creation and analyzing tool. The resulting network exhibited an average weighted degree of 23.041, a network diameter of 3 (indicating relatively short paths between any two technologies), and a graph density of 0.015. It showed a high propensity for local clustering with an average clustering coefficient of 0.912, and the average path length of 2.066. The entire filtered network comprised a single connected component. Community structure within this refined network was further investigated using two distinct algorithms. First, by applying the Louvain method a modularity score of 0.395 had been calculated. Second, the Girvan-Newman algorithm was also applied. This method identified 112 distinct communities and achieved a maximum modularity score of 0.44 (Fig~\ref{fig:full_girvan}). These metrics collectively describe the topological characteristics and the nuanced community organization of the refined technology co-occurrence network. These specific values and their implications for the BCI-XR field were detailed further in the Discussion section. the most prominent cluster based on Girvan-Newman consisted of 385 technologies, many of the most important technologies of the field (Highest weighted-degree) including BCI, VR, EEG, MI, Robotics were present in this cluster. Other notable cluster was the cluster including AR. The presence of AR in this cluster instead of the most prominent cluster is discussed further in Discussion. To visually explore these structural properties and the identified community structures, the final network representation of co-occurring technologies was then created using VOSviewer. This tool was employed for its capabilities in constructing and visualizing bibliometric maps based on co-occurrence data.

\section{Discussion}

This study introduced RATE, a novel multi-stage LLM-based pipeline for technology extraction, and applied it to map the technological landscape of research at the intersection of XR and BCI. The discussion first reflects on the RATE methodology, then interprets the thematic clusters and broader network structures identified, and finally considers the implications and future directions of this work.

\subsection{Performance and Methodological Contributions of RATE}

The RATE pipeline proved effective for technology extraction, achieving a high F1-score of 0.91 significantly outperforming state of the art supervised Machine Learning techniques such as BERT. Its success comes from an innovative multi-stage design. Initially, one LLM with RAG identifies a broad set of potential technology candidates. Then, crucially, a second LLM validates these candidates by assessing them against four distinct scholarly definitions of technology directly within their original textual context. This unique, definition-driven validation step significantly improves precision and helps overcome the common challenge of technology's ambiguous nature. RATE thus offers a robust approach for accurately mapping technological landscapes, reducing the reliance on extensive pre-annotated datasets often required by other methods. As a result, a novel method to detect technologies without requiring extensive field knowledge or annotated data had been proposed.

\begin{figure}[t]
			\centering
			\includegraphics[trim={0.5cm 5cm 0.5cm 5cm}, clip, scale = 0.4]{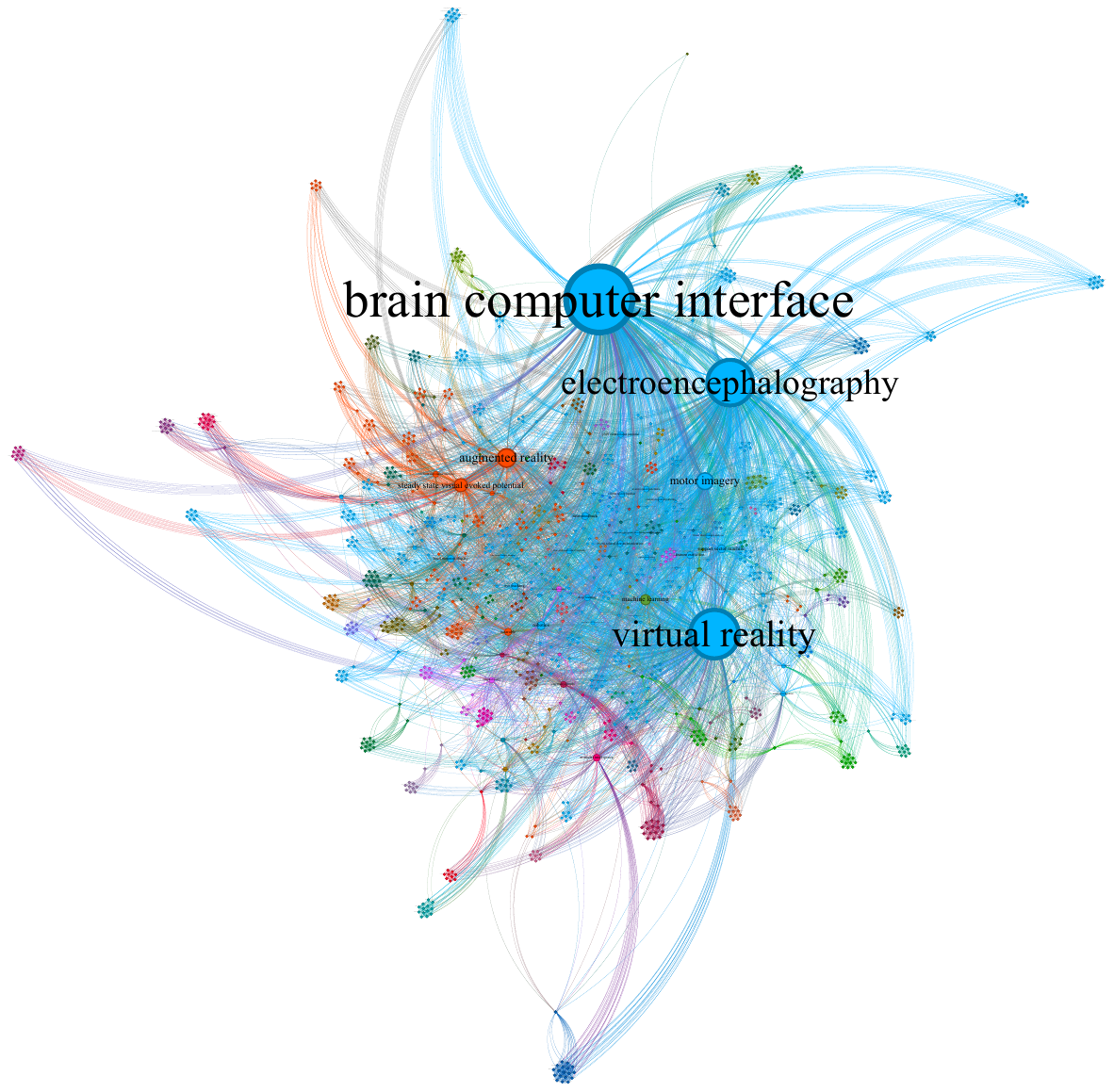}
				\caption{\small Representation of core technologies in Girvan-Newman communities}
				\label{fig:GIRVAN}
				
			\vspace{-1mm}
			\end{figure}

\begin{figure*}[t]
\vspace{-1em}
			\centering
			\includegraphics[width=185mm,scale=1.0]{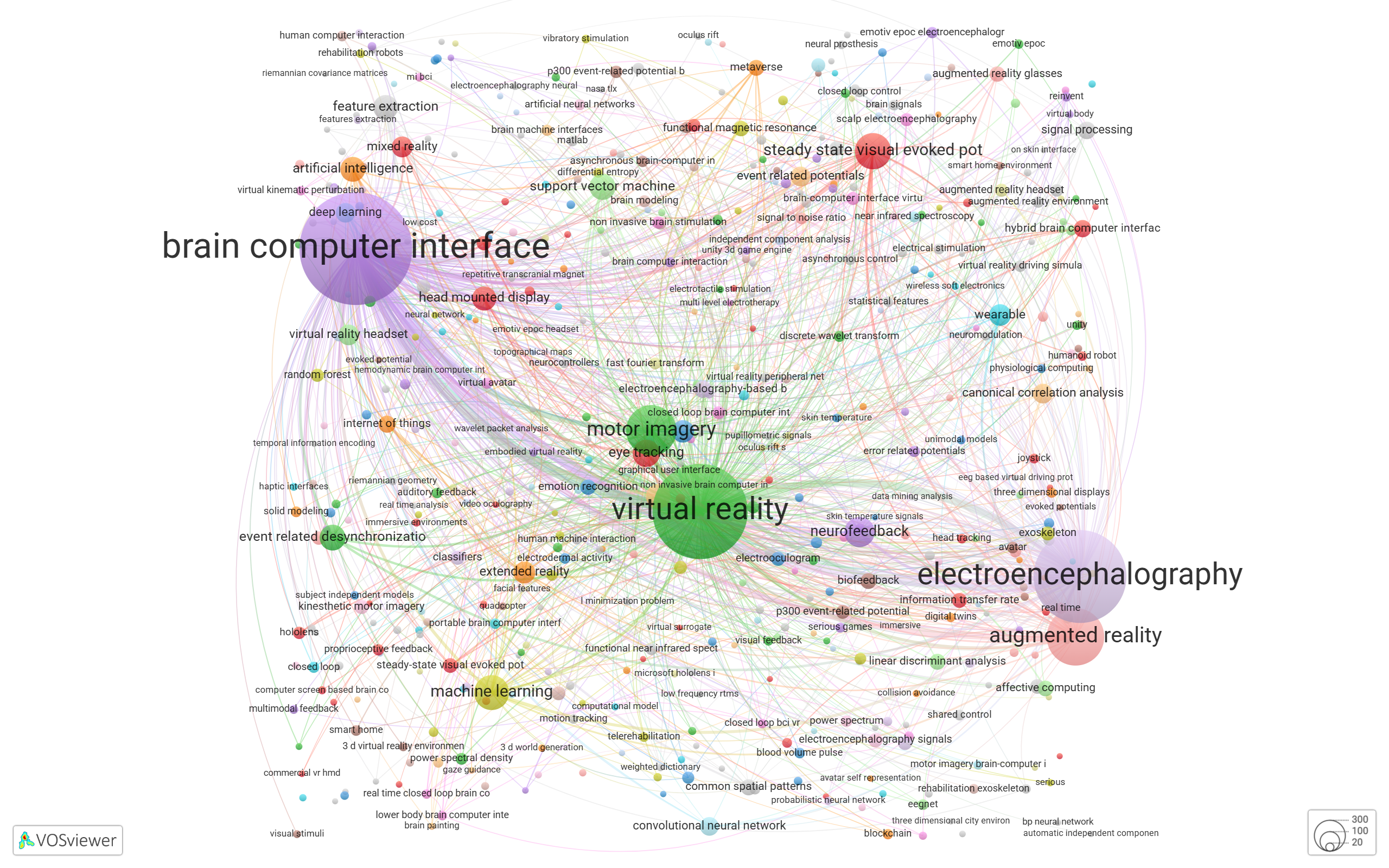}
				\caption{\small Representation of technologies co-occurrence network with VOSviewer}
				\label{fig:full}
			\end{figure*}

\subsection{Interpretation of Thematic Clusters in the BCI-XR Landscape}

The community detection analysis using the Girvan-Newman algorithm identified several key thematic clusters within the BCI-XR technology network (Fig~\ref{fig:GIRVAN}).
The most prominent cluster, was termed "\textit{VR-EEG Brain-Computer Interfaces and Neuro-rehabilitation Technologies}", shows the strong synergy between BCI, VR, and EEG related technologies. The significant presence of terms such as Rehabilitation technology, Exoskeletons, wheelchairs, motor priming, and Neurofeedback as distinct nodes within this core cluster highlights the primary application of integrated BCI-VR systems in neuro-rehabilitation. This finding, particularly the emphasis on VR over AR in medical and rehabilitation contexts, aligns with previous bibliometric analyses, such as Yeung et al. \cite{yeung2021virtual}. Furthermore, the inclusion of feature extraction, ML, and deep learning as significant nodes within this cluster points to the critical role of these computational methods in analyzing neural data and enabling BCI functionality within these rehabilitative applications.
A second distinct cluster, labeled "\textit{Portable AR-SSVEP Brain-Computer Interfaces for Real-Time Hands-Free Control}", centers around AR. This cluster's co-occurrence with technologies like portable systems, wearable devices, real-time processing, and hands-free control suggests a different application focus for BCI-AR integration compared to BCI-VR. The relative separation of AR from the main rehabilitation-focused cluster in the network indicates its potential development towards daily-use and interactive control applications, often using paradigms like SSVEP.
A third important cluster, named "\textit{Head-Mounted Haptic Virtual Reality Interfaces with Physiological Feedback}", groups HMD technologies with haptic feedback systems and various physiological measurement techniques, including hemodynamic BCIs, heart-rate monitoring, and broader brain wave detection methods. This suggests a research thrust focused on creating richer, multi-sensory immersive experiences where BCI inputs are potentially augmented or correlated with other physiological signals to enhance interaction or user state assessment.

\subsection{Broader Network Insights and Emerging Trends}

The overall network structure and the prominence of certain nodes reinforce the observation that VR, BCI, and EEG form a well-established core in this research domain, particularly within a rehabilitation context. While our data included mentions of non-EEG modalities like functional Near-Infrared Spectroscopy (fNIRS) and Magnetoencephalography (MEG), EEG clearly remains the dominant Neuroimaging technology in relation to BCI-XR applications found in the analyzed literature, likely due to its portability and cost-effectiveness.
The relative peripheral position or weaker linkage of terms like "\textit{Metaverse}" to core EEG or AI hubs in the network might suggest that, within this specific corpus and timeframe (2000-2024), its deep integration with foundational BCI technologies is still an emerging area rather than a central research theme. On the other hand, "machine learning" acts as a significant bridging node, connected to nearly all core technology areas, indicating its centrality and growing importance across the BCI-XR field. Our analysis indicates that Support Vector Machines (SVMs) were a frequently co-occurring ML method, often utilized for classification tasks based on extracted features. Convolutional Neural Networks (CNNs) also appeared as an important ML-based technology, particularly relevant for visual data processing in applications involving HMDs or BCI signal analysis.
Regarding emerging trends, while VR currently shows stronger and more established integration with BCI in this dataset (predominantly in rehabilitation), AR exhibits characteristics of a potent, growing area (Fig~\ref{fig:full}). Its distinct clustering with technologies emphasizing portability and real-time interaction (such as wearable systems and SSVEP-based BCI) suggests an evolving trajectory. Although currently less intertwined with the core BCI-rehabilitation cluster than VR, AR's role may expand significantly as portable BCI technology matures and new use cases beyond clinical settings become more prevalent.

\subsection{Future Directions}

Although the current framework does not incorporate extensive domain knowledge during the training phase, testing it across diverse domains and comparing evaluation metrics remains a critical task. Additionally, curating a comprehensive list of technologies to augment the RAG setup could substantially improve the model's precision. Therefore, future efforts should prioritize integrating such a thorough technology list into the model, followed by testing the framework in varied domains and a comparative analysis of the results.




\bibliographystyle{IEEEtran}
\bibliography{Biblo}

\begin{thebibliography}{10}
\providecommand{\url}[1]{#1}
\csname url@samestyle\endcsname
\providecommand{\newblock}{\relax}
\providecommand{\bibinfo}[2]{#2}
\providecommand{\BIBentrySTDinterwordspacing}{\spaceskip=0pt\relax}
\providecommand{\BIBentryALTinterwordstretchfactor}{4}
\providecommand{\BIBentryALTinterwordspacing}{\spaceskip=\fontdimen2\font plus
\BIBentryALTinterwordstretchfactor\fontdimen3\font minus \fontdimen4\font\relax}
\providecommand{\BIBforeignlanguage}[2]{{%
\expandafter\ifx\csname l@#1\endcsname\relax
\typeout{** WARNING: IEEEtran.bst: No hyphenation pattern has been}%
\typeout{** loaded for the language `#1'. Using the pattern for}%
\typeout{** the default language instead.}%
\else
\language=\csname l@#1\endcsname
\fi
#2}}
\providecommand{\BIBdecl}{\relax}
\BIBdecl

\bibitem{rauschnabel2022xr}
P.~A. Rauschnabel, R.~Felix, C.~Hinsch, H.~Shahab, and F.~Alt, ``What is xr? towards a framework for augmented and virtual reality,'' \emph{Computers in human behavior}, vol. 133, p. 107289, 2022.

\bibitem{dargan2023augmented}
S.~Dargan, S.~Bansal, M.~Kumar, A.~Mittal, and K.~Kumar, ``Augmented reality: A comprehensive review,'' \emph{Archives of Computational Methods in Engineering}, vol.~30, no.~2, pp. 1057--1080, 2023.

\bibitem{kohli2022review}
V.~Kohli, U.~Tripathi, V.~Chamola, B.~K. Rout, and S.~S. Kanhere, ``A review on virtual reality and augmented reality use-cases of brain computer interface based applications for smart cities,'' \emph{Microprocessors and Microsystems}, vol.~88, p. 104392, 2022.

\bibitem{yeung2021virtual}
A.~W.~K. Yeung, A.~Tosevska, E.~Klager, F.~Eibensteiner, D.~Laxar, J.~Stoyanov, M.~Glisic, S.~Zeiner, S.~T. Kulnik, R.~Crutzen \emph{et~al.}, ``Virtual and augmented reality applications in medicine: analysis of the scientific literature,'' \emph{Journal of medical internet research}, vol.~23, no.~2, p. e25499, 2021.

\bibitem{motejlek2021taxonomy}
J.~Motejlek and E.~Alpay, ``Taxonomy of virtual and augmented reality applications in education,'' \emph{IEEE transactions on learning technologies}, vol.~14, no.~3, pp. 415--429, 2021.

\bibitem{reljic2021augmented}
V.~Relji{\'c}, I.~Milenkovi{\'c}, S.~Dudi{\'c}, J.~{\v{S}}ulc, and B.~Baj{\v{c}}i, ``Augmented reality applications in industry 4.0 environment,'' \emph{Applied Sciences}, vol.~11, no.~12, p. 5592, 2021.

\bibitem{putze2020brain}
F.~Putze, A.~Vourvopoulos, A.~L{\'e}cuyer, D.~Krusienski, S.~Berm{\'u}dez~i Badia, T.~Mullen, and C.~Herff, ``Brain-computer interfaces and augmented/virtual reality,'' p. 144, 2020.

\bibitem{nicolas2012brain}
L.~F. Nicolas-Alonso and J.~Gomez-Gil, ``Brain computer interfaces, a review,'' \emph{sensors}, vol.~12, no.~2, pp. 1211--1279, 2012.

\bibitem{si2017brain}
H.~Si-Mohammed, F.~A. Sanz, G.~Casiez, N.~Roussel, and A.~L{\'e}cuyer, ``Brain-computer interfaces and augmented reality: A state of the art,'' in \emph{Graz Brain-Computer Interface Conference}, 2017.

\bibitem{piszcz2024impact}
A.~Piszcz, I.~Rojek, and D.~Miko{\l}ajewski, ``Impact of virtual reality on brain--computer interface performance in iot control—review of current state of knowledge,'' \emph{Applied Sciences}, vol.~14, no.~22, p. 10541, 2024.

\bibitem{sun2016mapping}
X.~Sun, K.~Ding, and Y.~Lin, ``Mapping the evolution of scientific fields based on cross-field authors,'' \emph{Journal of Informetrics}, vol.~10, no.~3, pp. 750--761, 2016.

\bibitem{arasti2010use}
M.~R. Arasti and N.~B. Moghaddam, ``Use of technology mapping in identification of fuel cell sub-technologies,'' \emph{international journal of hydrogen energy}, vol.~35, no.~17, pp. 9516--9525, 2010.

\bibitem{castells2000technology}
P.~E. Castells, M.~R. Salvador, and R.~M. Bosch, ``Technology mapping, business strategy, and market opportunities,'' \emph{Competitive Intelligence Review: Published in Cooperation with the Society of Competitive Intelligence Professionals}, vol.~11, no.~1, pp. 46--57, 2000.

\bibitem{firoozeh2020keyword}
N.~Firoozeh, A.~Nazarenko, F.~Alizon, and B.~Daille, ``Keyword extraction: Issues and methods,'' \emph{Natural Language Engineering}, vol.~26, no.~3, pp. 259--291, 2020.

\bibitem{siddiqi2015keyword}
S.~Siddiqi and A.~Sharan, ``Keyword and keyphrase extraction techniques: a literature review,'' \emph{International Journal of Computer Applications}, vol. 109, no.~2, 2015.

\bibitem{puccetti2023technology}
G.~Puccetti, V.~Giordano, I.~Spada, F.~Chiarello, and G.~Fantoni, ``Technology identification from patent texts: A novel named entity recognition method,'' \emph{Technological Forecasting and Social Change}, vol. 186, p. 122160, 2023.

\bibitem{minaee2025largelanguagemodelssurvey}
\BIBentryALTinterwordspacing
S.~Minaee, T.~Mikolov, N.~Nikzad, M.~Chenaghlu, R.~Socher, X.~Amatriain, and J.~Gao, ``Large language models: A survey,'' 2025. [Online]. Available: \url{https://arxiv.org/abs/2402.06196}
\BIBentrySTDinterwordspacing

\bibitem{zhao2023survey}
W.~X. Zhao, K.~Zhou, J.~Li, T.~Tang, X.~Wang, Y.~Hou, Y.~Min, B.~Zhang, J.~Zhang, Z.~Dong \emph{et~al.}, ``A survey of large language models,'' \emph{arXiv preprint arXiv:2303.18223}, vol.~1, no.~2, 2023.

\bibitem{gadetsky2025large}
A.~Gadetsky, A.~Atanov, Y.~Jiang, Z.~Gao, G.~H. Mighan, A.~Zamir, and M.~Brbic, ``Large (vision) language models are unsupervised in-context learners,'' \emph{arXiv preprint arXiv:2504.02349}, 2025.

\bibitem{gao2023retrieval}
Y.~Gao, Y.~Xiong, X.~Gao, K.~Jia, J.~Pan, Y.~Bi, Y.~Dai, J.~Sun, H.~Wang, and H.~Wang, ``Retrieval-augmented generation for large language models: A survey,'' \emph{arXiv preprint arXiv:2312.10997}, vol.~2, no.~1, 2023.

\bibitem{yin2023bibliometric}
Z.~Yin, Y.~Wan, H.~Fang, L.~Li, T.~Wang, Z.~Wang, and D.~Tan, ``Bibliometric analysis on brain-computer interfaces in a 30-year period,'' \emph{Applied Intelligence}, vol.~53, no.~12, pp. 16\,205--16\,225, 2023.

\bibitem{CSET2023}
\BIBentryALTinterwordspacing
{Center for Security and Emerging Technology}, ``Chinese catalogue of technologies prohibited or restricted from export,'' \url{https://cset.georgetown.edu/publication/china-export-control-catalog-2023}, Dec. 2023, (Center for Security and Emerging Technology, Trans.). Original work published December 21, 2023. [Online]. Available: \url{https://cset.georgetown.edu/publication/china-export-control-catalog-2023}
\BIBentrySTDinterwordspacing

\bibitem{IEA2025}
\BIBentryALTinterwordspacing
{International Energy Agency}, ``Etp clean energy technology guide,'' \url{https://www.iea.org/data-and-statistics/data-tools/etp-clean-energy-technology-guide}, Apr. 2025, (Online). [Online]. Available: \url{https://www.iea.org/data-and-statistics/data-tools/etp-clean-energy-technology-guide}
\BIBentrySTDinterwordspacing

\bibitem{ONET2024}
\BIBentryALTinterwordspacing
{National Center for O*NET Development}, ``O*net online,'' \url{https://www.onetonline.org/}, 2024, u.S. Department of Labor, Employment \& Training Administration. [Online]. Available: \url{https://www.onetonline.org/}
\BIBentrySTDinterwordspacing

\bibitem{Chase2024}
\BIBentryALTinterwordspacing
H.~Chase, ``Langchain (version 0.3.12) [computer software],'' \url{https://github.com/langchain-ai/langchain}, 2024, accessed: 2024-05-31. [Online]. Available: \url{https://github.com/langchain-ai/langchain}
\BIBentrySTDinterwordspacing

\bibitem{Ollama2025}
\BIBentryALTinterwordspacing
{Ollama}, ``Ollama (version 0.7.0) [computer software],'' \url{https://ollama.com/}, 2025, accessed: 2025-05-31. [Online]. Available: \url{https://ollama.com/}
\BIBentrySTDinterwordspacing

\bibitem{emb2024mxbai}
\BIBentryALTinterwordspacing
S.~Lee, A.~Shakir, D.~Koenig, and J.~Lipp. (2024) Open source strikes bread - new fluffy embedding model. [Online]. Available: \url{https://www.mixedbread.ai/blog/mxbai-embed-large-v1}
\BIBentrySTDinterwordspacing

\bibitem{liu2024deepseek}
A.~Liu, B.~Feng, B.~Xue, B.~Wang, B.~Wu, C.~Lu, C.~Zhao, C.~Deng, C.~Zhang, C.~Ruan \emph{et~al.}, ``Deepseek-v3 technical report,'' \emph{arXiv preprint arXiv:2412.19437}, 2024.

\bibitem{devlin2019bert}
J.~Devlin, M.-W. Chang, K.~Lee, and K.~Toutanova, ``Bert: Pre-training of deep bidirectional transformers for language understanding,'' in \emph{Proceedings of the 2019 conference of the North American chapter of the association for computational linguistics: human language technologies, volume 1 (long and short papers)}, 2019, pp. 4171--4186.

\bibitem{liu2021ner}
Z.~Liu, F.~Jiang, Y.~Hu, C.~Shi, and P.~Fung, ``Ner-bert: A pre-trained model for low-resource entity tagging,'' \emph{arXiv preprint arXiv:2112.00405}, 2021.

\bibitem{wolf2020huggingfacestransformersstateoftheartnatural}
\BIBentryALTinterwordspacing
T.~Wolf, L.~Debut, V.~Sanh, J.~Chaumond, C.~Delangue, A.~Moi, P.~Cistac, T.~Rault, R.~Louf, M.~Funtowicz, J.~Davison, S.~Shleifer, P.~von Platen, C.~Ma, Y.~Jernite, J.~Plu, C.~Xu, T.~L. Scao, S.~Gugger, M.~Drame, Q.~Lhoest, and A.~M. Rush, ``Huggingface's transformers: State-of-the-art natural language processing,'' 2020. [Online]. Available: \url{https://arxiv.org/abs/1910.03771}
\BIBentrySTDinterwordspacing

\bibitem{reback2020pandas}
\BIBentryALTinterwordspacing
T.~pandas~development team, ``pandas-dev/pandas: Pandas,'' Feb. 2020. [Online]. Available: \url{https://doi.org/10.5281/zenodo.3509134}
\BIBentrySTDinterwordspacing

\bibitem{hagberg2008exploring}
A.~Hagberg, P.~J. Swart, and D.~A. Schult, ``Exploring network structure, dynamics, and function using networkx,'' Los Alamos National Laboratory (LANL), Los Alamos, NM (United States), Tech. Rep., 2008.

\bibitem{blondel2008fast}
V.~D. Blondel, J.-L. Guillaume, R.~Lambiotte, and E.~Lefebvre, ``Fast unfolding of communities in large networks,'' \emph{Journal of statistical mechanics: theory and experiment}, vol. 2008, no.~10, p. P10008, 2008.

\bibitem{newman2004finding}
M.~E. Newman and M.~Girvan, ``Finding and evaluating community structure in networks,'' \emph{Physical review E}, vol.~69, no.~2, p. 026113, 2004.

\bibitem{bastian2009gephi}
M.~Bastian, S.~Heymann, and M.~Jacomy, ``Gephi: an open source software for exploring and manipulating networks,'' in \emph{Proceedings of the international AAAI conference on web and social media}, vol.~3, no.~1, 2009, pp. 361--362.

\bibitem{van2009software}
N.~Van~Eck and L.~Waltman, ``Software survey: Vosviewer, a computer program for bibliometric mapping,'' \emph{scientometrics}, vol.~84, no.~2, pp. 523--538, 2009.

\bibitem{li2025understanding}
M.~Li, H.~Chen, Y.~Wang, T.~Zhu, W.~Zhang, K.~Zhu, K.-F. Wong, and J.~Wang, ``Understanding and mitigating the bias inheritance in llm-based data augmentation on downstream tasks,'' \emph{arXiv preprint arXiv:2502.04419}, 2025.

\bibitem{ayoub2024inherent}
N.~F. Ayoub, K.~Balakrishnan, M.~S. Ayoub, T.~F. Barrett, A.~P. David, and S.~T. Gray, ``Inherent bias in large language models: A random sampling analysis,'' \emph{Mayo Clinic Proceedings: Digital Health}, vol.~2, no.~2, pp. 186--191, 2024.

\bibitem{pranckute2021web}
R.~Pranckut{\.e}, ``Web of science (wos) and scopus: The titans of bibliographic information in today’s academic world,'' \emph{Publications}, vol.~9, no.~1, p.~12, 2021.

\bibitem{agar2020technology}
J.~Agar, ``What is technology? technology: critical history of a concept, by eric schatzberg, chicago and london, university of chicago press, 2018, 352 pp., \$27.45 (paperback), isbn: 978-0-226-58383-9,'' \emph{Annals of Science}, vol.~77, no.~3, pp. 377--382, 2020.

\end{thebibliography}

\end{document}